\documentstyle[12pt]{article} 
\begin{document} \begin{titlepage}
\hspace{9cm} ULB--PMIF--92/02

\vspace{4.5cm} \begin{centering}

{\huge Comments on Unitarity in the Antifield Formalism}\\ \vspace{1cm}
{\large Glenn Barnich and Marc Henneaux$^*$\\ Facult\'e des Sciences,
Universit\'e Libre de Bruxelles,\\ Campus Plaine C.P. 231, B-1050
Bruxelles, Belgium}\\ \end{centering} \vspace{7cm}
{\footnotesize($^*$)Ma\^\i tre de Recherches au Fonds National de la
Recherche Scientifique. Also at Centro de Estudios Cient\'\i ficos de
Santiago, Chile.} \end{titlepage}

\begin{abstract} It is shown that the local completeness condition
introduced in the analysis of the locality of the gauge fixed action in the
antifield formalism plays also a key role in the proof of unitarity.
\end{abstract}
\pagebreak \section{Introduction} The antifield formalism is
a method to deal with the unphysical degrees of freedom of a gauge theory
in a manifestly covariant way.  However, the unitarity of the formalism is
not obvious without further investigation.

\noindent One indirect unitarity proof exists and is based on the following
argument :

(i) under quite general regularity conditions the antifield formalism is
equivalent to the Hamiltonian formulation of the BRST symmetry
\cite{mh1,mh2};

(ii) if the dynamics can be asymptotically linearized, the Kugo - Ojima
quartet mechanism \cite{ko} can be applied and the state cohomology of the
Hamiltonian BRST charge is isomorphic to the space of states in a physical
gauge.  \cite{mh3,as1,fs,as2}. It is thus endowed with a positive definite
inner product \footnote{We assume, of course, that the physical (i.e.,
gauge invariant) degrees of freedom are quantized with a positive metric.}.

More recently, a different approach has been taken, which investigates
unitarity directly at the Lagrangian level without going through the
Hamiltonian BRST formulation \cite{sf,as1,fs,as2,jg}. This approach also
proceeds within the framework of perturbation theory and considers the
asymptotic, (i.e., linearized) L-th order reducible gauge theory described
by

- a quadratic action \begin{equation} S^{(0)}=\int dt\ L
(\phi^i,\dot\phi^i) \end{equation} depending on the fields and their first
order derivatives

- reducible gauge generators of the form \begin{equation}
\delta_{\epsilon}\phi^i = R^i_{(0)0\alpha}\epsilon^\alpha +
R^i_{(0)1\alpha}\dot\epsilon^\alpha \end{equation}

- reducibility coefficients of the form \begin{equation}
R^{\alpha_k}_{(k)\alpha_{k+1}}(t,t^\prime )=
R^{\alpha_k}_{(k)0\alpha_{k+1}}\delta(t-t^\prime )
+R^{\alpha_k}_{(k)1\alpha_{k+1}}{d\over dt}\delta (t-t^\prime ) \qquad
k=1,...,L \end{equation} where the $R^{\alpha_k}_{(k)l\alpha_{k+1}}\
^\prime$s $(l=0,1\quad k=0,...,L\quad \alpha_0 =i\quad \alpha_1=\alpha)$
are all constant c-number matrices.

\noindent The authors of \cite{sf} conclude that unitarity holds if and
only if $R_{(L)0}$ is of maximum rank \begin{equation} rank\
R^{\alpha_L}_{(L)0\alpha_{L+1}} = m_{L+1},\quad \alpha_{L+1}=1,...,m_{L+1}
\end{equation} and the $R_{(k)0}\ ^\prime$ s "form an exact sequence"
meaning that \begin{eqnarray} \lefteqn{Im\ R_{(k)0} \equiv \lbrace
A^{\alpha_k} = R^{\alpha_k}_{(k)0\alpha_{k+1}} \lambda^{\alpha_{k+1}},\
\lambda^{\alpha_{k+1}}\ arbitrary \rbrace }\nonumber \\ & = Ker\ R_{(k-1)0}
\equiv \lbrace A^{\alpha_k} \vert R^{\alpha_{k-1}}_{(k-1)0\alpha_{k}}
A^{\alpha_k} = 0 \rbrace\ .  \end{eqnarray}

\noindent \lbrack The extra conditions on the $R_{(k)1}\ ^\prime$s imposed
in \cite{sf} are actually simplifying assumptions to prevent the appearance
of second class constraints in the passage from the gauge fixed Lagrangian
action to the canonical formalism and are thus not essential for
unitarity.\rbrack

But unitarity of the theory is guaranteed to hold since the general
analysis based on the Hamiltonian formalism applies. Indeed, the regularity
conditions are fulfilled for a quadratic action, because all relevant
matrices are constant and have thus constant ranks. This means that one
should be able to prove $(4)$ and $(5)$ by using properties of the
Hamiltonian formulation.

It is however of interest to see how the conditions $(4)$ and $(5)$ hold
independently of Hamiltonian arguments. This letter provides a direct
Lagrangian proof that the conditions $(4)$ and $(5)$ are automatically
fulfilled - and so need not be imposed as extra conditions. This proof
relies on the fact that the gauge transformations and the reducibility
coefficients should be chosen so as to be not only complete but also
locally complete \cite{mh4}.  Hence, the local completeness requirement,
which was first introduced in order to control locality of the gauge fixed
action, plays also a key role in the proof of unitarity.  This is not
surprising since locality (in time) and unitarity are known to be
intimately connected.  \section{Local completeness} We will consider the
general case in wich the free action can depend on the fields and their
derivatives up to some finite order that may be higher than one :
\begin{equation} S^{(0)}=\int dt\ L (\phi^i,\dot\phi^i,...,\phi^{i(k)})\ .
\end{equation} For notational simplicity we assume all the classical fields
to be bosonic.

\noindent Since the action is quadratic, gauge transformations do not
depend on the fields and take the general form \begin{equation}
\delta\phi^i(t) = \int dt^\prime\
R^i_{(0)\alpha}(t,t^\prime)\epsilon^\alpha(t^\prime) \end{equation} with
\begin{equation} R^i_{(0)\alpha}(t,t^\prime)=\sum_{l=0}^m
R^i_{(0)l\alpha}{d^l\over dt^l}\delta(t-t^\prime) \end{equation} for some
finite $m$. The $R_{(0)l}\ ^\prime$s are c-number matrices.

As explained in \cite{mh4}, the gauge generators and the reducibility
functions should be taken to be locally complete. Let us briefly review
what this means in the context of the linear theory based on $(6)$.
\subsection{Definition} The gauge generators $(7)$ are locally complete in
time if any transformation \begin{equation} \delta\phi^i(t) = \int
dt^\prime\ a^i(t,t^\prime)u(t^\prime),\quad
a^i(t,t^\prime)=\sum_{l=0}^{m_a} a^i_l{d^l\over dt^l}\delta(t-t^\prime)
\end{equation} leaving the action invariant (up to a boundary term) for any
choice of the arbitrary function u(t), corresponds to $(7)$ (up to gauge
transformations vanishing on-shell) with a special choice of the gauge
parameter ("completeness") involving the arbitrary function u(t) and a
finite number of its time derivatives ("local completeness", i.e.,
"locality in time of the gauge transformations"). That is, one must be able
to reproduce $(9)$ by choosing $\epsilon^\alpha$ in $(7)$ to be of the
following form \begin{equation} \epsilon^\alpha(t)=\int dt^\prime\ w^\alpha
(t,t^\prime)u(t^\prime),\quad w^\alpha (t,t^\prime)=\sum_{l=0}^{m_w}
w^i_l{d^l\over dt^l}\delta(t-t^\prime)\ .  \end{equation} So there is no
gauge transformation that is not included among $(7)$.

\noindent The fact that $(9)$ is included among $(7)$ implies
\begin{eqnarray} \lefteqn{a^i(t,t^\prime)=\int dt^{\prime\prime}\
R^i_{(0)\alpha}(t,t^{\prime\prime}) w^\alpha
(t^{\prime\prime},t^\prime)}\nonumber \\ & \Longleftrightarrow\
a^i_l=\displaystyle{ \sum_{j=0}^{l}} R^i_{(0)l-j\alpha}w^\alpha_j,\quad
l=0,...,m_a+m_w \end{eqnarray} with the convention : $a^i_k=0\quad
k>m_a,\quad R^i_{(0)k\alpha}=0\quad k>m$,

\noindent $w^\alpha_k=0\quad k>m_w$.  \subsection{Equivalent formulation of
local completeness} Because $(7)$ are gauge transformations, one has the
Noether identities \begin{equation} \sum_{k=0}^{m}{d^k\over
dt^k}(R^i_{(0)k\alpha}{\delta S\over\delta\phi^i})=0\ .  \end{equation}
Now, the transformation $(9)$ leaves the action invariant if and only if
\begin{equation} \sum_{k=0}^{m_a}{d^k\over dt^k}(a^i_{k}{\delta
S\over\delta\phi^i})=0\ .  \end{equation} Hence, the gauge transformations
are locally complete in time if and only if the holding of $(13)$ for
$a^i_k$ implies that $a^i_k$ can be written as in $(11)$.  That is $(12)$
"should exhaust all the identities".  \subsection{Higher order reducibility
coefficients} The concept of local completeness for the higher order
reducibility coefficients is defined in a similar manner.  At each stage,
the reducibility coefficients \begin{equation}
R^{\alpha_k}_{(k)\alpha_{k+1}}
(t,t^\prime)=\sum_{l=0}^{m_k}R^{\alpha_k}_{(k)l\alpha_{k+1}} {d^l\over
dt^l}\delta (t-t^\prime) \end{equation} must be such that \begin{equation}
\int dt^\prime\
R^{\alpha_{k-1}}_{(k-1)\alpha_{k}}(t,t^\prime)R^{\alpha_k}_{(k)\alpha_{k+1}}
(t^\prime,t^{\prime\prime})=0 \end{equation} and if \begin{equation} \int
dt^\prime\
R^{\alpha_{k-1}}_{(k-1)\alpha_{k}}(t,t^\prime)v^{\alpha_k}(t^\prime)=0
\end{equation} where $v^{\alpha_k}(t)$ is of the form \begin{equation}
v^{\alpha_k}(t)=\sum_{l=0}^{m_v}v^{\alpha_k}_l{d^l\over dt^l}u(t)
\end{equation} and involves the arbitrary function u(t) and a finite number
of its time derivatives, then \begin{equation} v^{\alpha_k}(t) =\int
dt^\prime\
R^{\alpha_k}_{(k)\alpha_{k+1}}(t,t^\prime)w^{\alpha_{k+1}}(t^\prime)
\end{equation} for some $w^{\alpha_{k+1}}(t)$ (completeness), where the
$w^{\alpha_{k+1}}(t)$ are functions of u(t) and a finite number of its time
derivatives (local completeness, i.e., "locality in time of reducibility")
\begin{equation}
w^{\alpha_{k+1}}(t)=\sum_{l=0}^{m_w}w^{\alpha_{k+1}}_l{d^l\over dt^l}u(t)\
.  \end{equation} In terms of the matrices
$R^{\alpha_{k-1}}_{(k-1)l\alpha_{k}},\ R^{\alpha_{k}}_{(k)l\alpha_{k+1}}\ ,
v^{\alpha_k}_l,\ w^{\alpha_{k+1}}_l$ the equation $(15)$ becomes
\begin{equation} \sum_{j=0}^{n}R^{\alpha_{k-1}}_{(k-1)n-j\alpha_{k}}
R^{\alpha_{k}}_{(k)j\alpha_{k+1}}=0,\quad n=0,...,m_{k-1}+m_k
\end{equation} where $R^{\alpha_{k-1}}_{(k-1)n-j\alpha_{k}}=0\quad
n-j>m_{k-1},\quad R^{\alpha_{k}}_{(k)j\alpha_{k+1}}=0\quad j>m_k$.
Furthermore, if \begin{equation}
\sum_{j=0}^{n^\prime}R^{\alpha_{k-1}}_{(k-1)n^\prime-j\alpha_{k}}
v^{\alpha_k}_j=0,\quad n^\prime=0,...,m_{k-1}+m_v \end{equation} where
$v^{\alpha_k}_j=0\quad j>m_v$, then one must have \begin{equation}
v^{\alpha_k}_l =\sum_{j=0}^{l} R^{\alpha_{k}}_{(k)l-j\alpha_{k+1}}
w^{\alpha_{k+1}}_j, \quad l=0,...,m_k+m_w \end{equation} with
$w^{\alpha_{k+1}}_j=0\quad j>m_w$ and $m_k+m_w\geq m_v$.
\subsection{Remarks} (i) The gauge transformations and the reducibility
coefficients should always be taken that way. This can always be done.
However, the reducibility of the theory may be infinite if further
requirements (such as Lorentz covariance) are added.  An example where this
situation arises is given by the N=1 superparticle \cite{bs}.

\noindent (ii) The equations $(21)$ are symmetrical between $n^\prime=0$
and $n^\prime=m_{k-1}+m_v$.  This symmetry is only superficial, however,
because one can introduce higher order coefficients $v^{\alpha_k}_j$
($j>m_v$) by adding to $v^{\alpha_k}(t)$ time derivatives of the
reducibility identities. This modifies the equation $(21)$ with
$n^\prime=m_{k-1}+m_v$, which is no longer the last equation, but leaves
$(21)$ with $n^\prime=0$ unchanged.  \section{Example} Let us consider the
pure gauge theory with action $S(\phi)=0$.  The transformation $\delta
\phi=\dot\epsilon,\ \epsilon=\epsilon (t)$ leaves the action invariant.
But it is not locally complete because the transformation
$\delta\phi=\eta,\ \eta=\ \eta(t)$ also leaves the action invariant, but
one cannot obtain $\delta\phi=\eta$ from $\delta\phi=\dot\epsilon$ by a
choice of $\epsilon =k_0 \eta +k_1 \dot\eta +..+k_m\eta^{(m)}$.  (To
eliminate the derivative, one needs to integrate and consider the boundary
conditions, wich is not a local, algebraic manipulation.)  By contrast, the
transformation $\delta\phi=\eta$ is locally complete.  So,
$\delta\phi=\dot\epsilon$ alone is an inappropriate starting point, but
$\delta\phi=\eta$ is all right.

\noindent A solution to the master equation is given in this simple example
by \begin{equation} S=\int dt\ \phi^*RC + \bar C^*\bar\pi \end{equation}
where $\bar C$ and $\bar\pi$ are variables of the non-minimal sector.
Choosing the gauge-fixing fermion as \begin{equation} \psi=\int dt\ \bar
C(\omega\phi+{1\over 2}\bar\pi) \end{equation} with
$\omega=\alpha+\beta{d\over dt}+...+\nu{d^k\over dt^k}$,
$\alpha,\beta,...,\nu$ constant, the gauged fixed action is
\begin{equation} S_\psi=\int dt\ \bar C\omega RC +
\bar\pi\omega\phi+{1\over 2}\bar\pi^2\ .  \end{equation}

- Taking $\omega=1$ and eliminating the auxiliary variable $\bar\pi$ by
means of its equation of motion, we get, for the locally complete form of
the gauge transformation ($R=\delta(t-t^\prime)$) \begin{equation}
S_{\psi_R}=\int dt\ \bar CC -{1\over 2}\phi^2\ .  \end{equation} The
primary constraint $\ p_{\bar C}\approx 0,\ p_C\approx0,\ p_\phi\approx0$
induce the secondary constraints $C\approx0,\ \bar C\approx0,\
\phi\approx0$ and all the constaints are second class. The Dirac bracket
method then eliminates the pure gauge variable $\phi$, the ghost and
antighost as well as their respective momenta. The theory is trivially
unitary since there is only one physical state.  The occurence of second
class constraints folows from the fact that $R_{(0)1}=0$.

-With $R={d\over dt}\delta(t-t^\prime)$ however (and $\omega$ still equal
to $1$), \begin{equation} \tilde S_{\psi_R}=\int dt\ \bar C\dot C -{1\over
2}\phi^2\ .  \end{equation} The primary constraints $p_C+\bar C\approx 0,\
p_{\bar C}\approx 0,p_\phi\approx0$ induce the secondary constraint
$\phi\approx0$. These four constraints are second class, and using the
Dirac bracket, one can set $\bar C=-p_C,\ p_{\bar C}=p_\phi=\phi=0$ in the
action \begin{equation} S_H=\int dt\ p_C\dot C+p_{\bar C}\dot{\bar C} +
p_\phi\dot\phi-{1\over 2}\bar\phi^2+\lambda (p_C+\bar C) +\mu p_{\bar
C}+\nu p_\phi \end{equation} getting \begin{equation} S_{H\prime}=\int dt\
p_C\dot C\ .  \end{equation}

\noindent The BRST symmetry for $\tilde S_{\psi_R}$ is given by
$\delta_\epsilon\bar C=-\phi\epsilon,\ \delta_\epsilon\phi=\dot C\epsilon$
and the corresponding Noether charge expressed in phase space reads
$\Omega=p_{\bar C}\phi+p_\phi\dot C$ and vanishes when enforcing the second
class constraints : $\Omega^\prime=0$. So $\Omega^\prime$ acts neither on
$C$ nor on $p_C$. These variables remain in the BRST cohomology and the
theory based on the incorrect description $(27)$ contains extra states
besides the physical state. As a result, it is not unitary.  \section{Proof
of $(4)$ and $(5)$} We will assume from now on that the local completeness
conditions are fulfilled.  \subsection{Theorem} (i) $R_{(L)0}$ is of
maximal rank

\noindent (ii) the $R_{(k)0}\ ^\prime$s form an exact sequence.

\noindent The proof of the theorem relies on the following lemmas :
\subsection{Lemma 1} If $R^i_{(0)0\alpha}$ is not of maximum rank, the
gauge transformations are reducible.  The system is then an L-th stage
reducible theory with L$\geq$1, and the 0-th stage cannot be the last step.

\noindent Proof : One has \begin{equation} R^i_{(0)0\alpha}\mu^\alpha_a=0
\end{equation} for some $\mu^\alpha_a\neq 0$ that are constant vectors.
Thus from $(12)$ one gets \begin{equation} \sum_{k=1}^{m}{d^k\over
dt^k}(R^i_{(0)k\alpha}{\delta S\over\delta\phi^i}\mu^\alpha_a)=0\qquad
(identically) \end{equation} and so \begin{equation}
\sum_{k=0}^{m}{d^{k-1}\over dt^{k-1}}(R^i_{(0)k\alpha}{\delta
S\over\delta\phi^i}\mu^\alpha_a)=C\qquad (identically) \end{equation} where
$C$ is a constant independent of the fields.

\noindent But the left-hand side vanishes on-shell, which implies $C=0$,
i.e., \begin{equation} \sum_{k=0}^{m-1}{d^k\over
dt^k}(R^i_{(0)k+1\alpha}{\delta S\over\delta\phi^i}\mu^\alpha_a)=0,
\end{equation} which is of the form $(13)$.  According to the local
completeness assumption there exists $w^\alpha_{la},\ l=0,...,m_a$ such
that \begin{equation}
R^i_{(0)k\alpha}\mu^\alpha_a=\sum_{j=0}^{k-1}R^i_{(0)k-1-j\alpha}w^\alpha_{ja},
\quad k=1,...,m+m_a+1\ .  \end{equation} Defining
$v^\alpha_{0a}=\mu^\alpha_a,\ v^\alpha_{j+1a}=-w^\alpha_{ja}$ and using
$(30)$ this gives : the systems \begin{equation}
\sum_{j=0}^{k}R^i_{(0)k-j\alpha}v^\alpha_{ja}=0,\quad k=1,...,m+m_a+1
\end{equation} possess non trivial solutions for each a.  This implies that
the gauge transformations defined by $R^i_{(0)\alpha}(t,t^\prime)$ are
reducible because \begin{equation} (35)\Longleftrightarrow\int dt^\prime\
R^i_{(0)\alpha}v^\alpha_{a}(t^\prime,t^{\prime\prime}) =0 \end{equation}
where \begin{equation}
v^\alpha_{a}(t,t^\prime)=\sum_{l=0}^{m_a+1}v^\alpha_{la}{d^l\over
dt^l}\delta(t-t^\prime) \end{equation} or in other words, the non-vanishing
gauge functions \begin{equation} \epsilon^\alpha_{a}(t)=\int dt^\prime\
v^\alpha_{a}(t,t^\prime)u(t^\prime) \end{equation} yield $\delta\phi^i=0$.
\subsection{Lemma 2} If $R^{\alpha_k}_{(k)0\alpha_{k+1}}$ has zero (right)
eigenvectors \begin{equation}
R^{\alpha_k}_{(k)0\alpha_{k+1}}\mu^{\alpha_{k+1}}_a
=0\qquad\mu^{\alpha_{k+1}}_a\neq0
\end{equation} then the $R_{(k)}\ ^\prime$s are reducible. More precisely,
to each zero eigenvector of the $R_{(k)0}\ ^\prime$s corresponds one
reducibility identity.

\noindent Proof : The proof proceeds in a similar way. Consider the vector
\begin{equation}
v^{\alpha_k}_{la}=R^{\alpha_k}_{(k)l+1\alpha_{k+1}}\mu^{\alpha_{k+1}}_a\ .
\end{equation} Then \begin{equation}
\sum_{j=0}^{n}R^{\alpha_{k-1}}_{(k-1)n-j\alpha_{k}}v^{\alpha_k}_{ja}=0,\quad
n=0,...,m_{k-1}+m_k-1\ .  \end{equation} Indeed using $(40)$ and $(20)$, we
get for the left-hand side of $(41)$ \begin{equation}
\sum_{j=0}^{n}R^{\alpha_{k-1}}_{(k-1)n-j\alpha_{k}}
R^{\alpha_k}_{(k)j+1\alpha_{k+1}}
\mu^{\alpha_{k+1}}_a=-R^{\alpha_{k-1}}_{(k-1)n+1\alpha_{k}}R^
{\alpha_k}_{(k)0\alpha_{k+1}} \mu^{\alpha_{k+1}}_a=0\ .  \end{equation} The
local completeness assumption on
$R^{\alpha_k}_{(k)\alpha_{k+1}}(t,t^\prime)$ implies the existence of
$w^{\alpha_{k+1}}_{ja}$ such that $(22)$ is satisfied for our choice of
$v^{\alpha_k}_{ja}$ : \begin{equation}
R^{\alpha_k}_{(k)l\alpha_{k+1}}\mu^{\alpha_{k+1}}_a=\sum_{j=0}^{l-1}
R^{\alpha_k}_{(k)l-j\alpha_{k+1}}w^{\alpha_{k+1}}_{ja},\quad
l=1,...,m_k+1+m_w\ .  \end{equation} Together with
$R^{\alpha_k}_{(k)0\alpha_{k+1}}\mu^{\alpha_{k+1}}_a=0$, we find
\begin{equation} \sum_{j=0}^{n}
R^{\alpha_k}_{(k)n-j\alpha_{k+1}}v^{\alpha_{k+1}}_{ja}=0
\end{equation} where $v^{\alpha_{k+1}}_{0a}=\mu^{\alpha_{k+1}}_a,\
v^{\alpha_{k+1}}_{j+1a}=-w^{\alpha_{k+1}}_{ja}$.  This implies
that the reducibility coefficients
$R^{\alpha_k}_{(k)\alpha_{k+1}}(t,t^\prime)$ are reducible,
\begin{equation} \int dt^\prime\
R^{\alpha_k}_{(k)\alpha_{k+1}}(t,t^\prime)v^{\alpha_{k+1}}_{a}(t^\prime,
t^{\prime\prime})=0 \end{equation} where \begin{equation}
v^{\alpha_{k+1}}_{a}(t,t^\prime)=
\sum_{l=0}^{m_w+1}v^{\alpha_{k+1}}_{la}{d^l\over dt^l}
\delta(t-t^\prime)\ .
\end{equation} We are now in a position to prove the theorem.
\subsection{Proof of the theorem} (i) If $R_{(L)0}$ had a right zero
eigenvector, then, according to the lemmas, the $R_{(L)}\ ^\prime$s would
be reducible and the theory, contrary to the initial hypothesis, cannot be
a L-th stage reducible theory. Hence, $R_{(L)0}$ is of maximum rank.

\noindent (ii) $\int dt^\prime\
R^{\alpha_{k-1}}_{(k-1)\alpha_{k}}(t,t^\prime)R^{\alpha_k}_{(k)\alpha_{k+1}}
(t^\prime,t^{\prime\prime})=0$ implies at order zero that

\noindent $R^{\alpha_{k-1}}_{(k-1)0\alpha_{k}}
R^{\alpha_k}_{(k)0\alpha_{k+1}}=0$ and hence that $Im\ R_{(k)0}\subset Ker\
R_{(k-1)0}$.

\noindent Conversely, let us assume that
$R^{\alpha_{k-1}}_{(k-1)0\alpha_{k}}v^{\alpha_k}_0=0$.  Then, by the second
lemma, there is a reducibility identity associated with $v^{\alpha_k}_0$,
i.e., there exists $v^{\alpha_k}(t)=v^{\alpha_k}_0\delta (t-t^\prime)+...$
such that \begin{equation} \int dt^\prime\
R^{\alpha_{k-1}}_{(k-1)\alpha_{k}}(t,t^\prime)v^{\alpha_k}(t^\prime)=0\ .
\end{equation} The local completeness assumption $(16)-(19)$ implies then
\begin{equation} v^{\alpha_k}(t)= \int dt^\prime\
R^{\alpha_{k}}_{(k)\alpha_{k+1}}(t,t^\prime)w^{\alpha_{k+1}}(t^\prime)
\end{equation} which reads, at order zero, \begin{equation}
v^{\alpha_k}_0=R^{\alpha_{k}}_{(k)0\alpha_{k+1}}w^{\alpha_{k+1}}_0\ .
\end{equation} Consequently $Ker\ R_{(k-1)0}\subset Im\ R_{(k)0}$.
\subsection{Remarks} (i) If only $R^{\alpha_{k}}_{(k)0\alpha_{k+1}}$ and
$R^{\alpha_{k}}_{(k)1\alpha_{k+1}}$ are different from zero,

\noindent $rank\ R^{\alpha_{k}}_{(k)1\alpha_{k+1}}\leq rank\
R^{\alpha_{k}}_{(k)0\alpha_{k+1}}$. Indeed, the systems $(35)$ or $(44)$
have non trivial solutions for each $a$. Suppose then that $\lambda^a_b
R^{\alpha_k}_{(k)1\alpha_{k+1}} v^{\alpha_{k+1}}_{0a}=0,\quad b=1,...,B$
define a complete and independent set of such relations.  Replacing $B$ of
the $v^{\alpha_{k+1}}_{0a}\ ^\prime$s by $\tilde v^{\alpha_{k+1}}_{0b}=
\lambda^a_bv^{\alpha_{k+1}}_{0a}$ in such a way that the new set $\tilde
v^{\alpha_{k+1}}_{0a}$ is still a complete set of zero eigenvectors of $
R^{\alpha_{k}}_{(k)0\alpha_{k+1}}$, we find that the $\tilde
v^{\alpha_{k+1}}_{0b}$ are also zero eigenvectors of
$R^{\alpha_{k}}_{(k)1\alpha_{k+1}}$. The remaining $\tilde
v^{\alpha_{k+1}}_{0c}$ give rise to linearly independent $\tilde
v^{\alpha_{k+1}}_{1c}\ ^\prime$s. The next step consists in discussing in
the same way the linear dependence of $R^{\alpha_{k}}_{(k)1\alpha_{k+1}}
\tilde v^{\alpha_{k+1}}_{1c}$. At the end of the redefinitions, one finds
that to independent zero eigenvectors of
$R^{\alpha_{k}}_{(k)0\alpha_{k+1}}$ correspond independent zero
eigenvectors of $R^{\alpha_{k}}_{(k)1\alpha_{k+1}}$ which implies the
stated rank condition.

\noindent (ii) The condition considered in \cite{jg} that
$R^{\alpha_{k}}_{(k)1\alpha_{k+1}} \neq
N_{\alpha_{k+1}}^{\beta_{k+1}}R^{\alpha_{k}}_{(k)0\beta_{k+1}}$ with
$N_{\alpha_{k+1}}^{\beta_{k+1}}$ invertible matrices also follows from the
local completeness assumption. Indeed suppose that
$R^{\alpha_{k}}_{(k)1\alpha_{k+1}}
=N_{\alpha_{k+1}}^{\beta_{k+1}}R^{\alpha_{k}}_{(k)0\beta_{k+1}}$.  The
reducibility functions then reads \begin{equation}
R^{\alpha_{k}}_{(k)\alpha_{k+1}}(t,t^\prime)=
R^{\alpha_{k}}_{(k)0\alpha_{k+1}}\delta(t-t^\prime)
+N_{\alpha_{k+1}}^{\beta_{k+1}}R^{\alpha_{k}}_{(k)0\beta_{k+1}}{d\over
dt}\delta(t-t^\prime) \end{equation} and is not locally complete because
\begin{equation} \tilde
R^{\alpha_{k}}_{(k)\alpha_{k+1}}(t,t^\prime)=R^{\alpha_{k}}_{(k)0\alpha_{k+1}}
\delta(t-t^\prime) \end{equation} is also a k-th order reducibility
function.  Indeed, writing the system $(20)$ for the particular
reducibility function $(50)$, implies the system $(20)$ corresponding to
the reducibility function $(51)$.  However, trying to express $(51)$ in
terms of $(50)$ as in $(18)$ or $(22)$ requires an infinite number of
matrices $w^{\alpha_{k+1}}_{l\beta_{k+1}}$.  \section{Conclusion} The local
completeness assumption on the gauge generators and the reducibility
coefficients has been introduced in \cite{mh4} in the study of the locality
of the gauge fixed action.  We have shown here that it plays also a crucial
role in guaranteeing unitarity of the antifield formalism.
\section{Acknowledgments} One of us (M.H.) is grateful to A.A. Slavnov for
useful conversations. This work has been supported in part by a research
contract with the Commission of the European Communities.  \pagebreak

\end{document}